\documentclass{ws-procs9x6}

\begin{document}

\title{
The chiral limit in lattice QCD
}

\author{Hidenori Fukaya (for JLQCD collaboration)}

\address{
Theoretical Physics Laboratory, RIKEN\\
Wako, Saitama 351-0198, Japan\\
E-mail: hfukaya@riken.jp
}

\begin{abstract}
It has been a big challenge for lattice QCD to 
simulate dynamical quarks near the chiral limit.
Theoretically, it is well-known that the naive chiral
symmetry cannot be realized on the lattice (the
Nielsen-Ninomiya theorem).
Also practically, the computational cost rapidly grows as
the quark mass is reduced. 
The JLQCD collaboration started a project to perform
simulations with exact but modified chiral symmetry using
the the overlap-Dirac operator and the topology conserving action.
The latter is helpful to reduce the numerical cost of 
the dynamical quarks.
Our simulation of two-flavor QCD has been successful to
reduce the sea quark mass down to a few MeV. 
\end{abstract}


\bodymatter

\section{Introduction}\label{aba:sec1}

Lattice QCD, has
been successful to analyze the low-energy dynamics of mesons
and baryons. 
Non-perturbative quantities, such as the hadron spectrum,
matrix elements,  the chiral phase transition {\it etc.}
have been investigated by large-scale calculations often
using supercomputers. 

The lattice regularization, however, violates some important
symmetries that the continuum theory has.
For instance, the translational invariance is violated
except for its discrete subgroup. 
The chiral symmetry, one of the most important symmetries of
QCD, is also difficult to realize on the lattice. 
This is known as the fermion-doubling problem that 
any chiral Dirac operator must have unphysical poles
\cite{Nielsen:1980rz, Nielsen:1981xu}.
In order to eliminate the doubler modes, one has to give up
the chiral symmetry, or to employ complicated Dirac
operators satisfying the Ginsparg-Wilson relation
\cite{Kaplan:1992bt, Shamir:1993zy,
Furman:1994ky,Neuberger:1997fp, Neuberger:1998wv,Ginsparg:1981bj}, 
for which the locality is less obvious and more importantly
much larger numerical cost is required compared to the
explicitly local (the so-called ultra-local) Dirac operators.
Most of the previous QCD simulations with the dynamical quarks
were, therefore, limited to those with the Dirac operator which 
explicitly breaks the chiral symmetry.
Moreover, the sea quark mass in such simulations were much
larger than the physical values.
Their results may contain systematic effects due to additive
quark mass renormalizaiton, unwanted operator mixings with
opposite chirality, additional symmetry breaking terms in
the chiral perturbation theory, chiral extrapolation from
rather heavy quark masses, and so on.

The JLQCD collaboration started a new project to simulate
QCD near the chiral limit.
It uses a new supercomputer system installed at KEK in early
2006. 
We employ the overlap-Dirac operator 
\cite{Neuberger:1997fp, Neuberger:1998wv},
which satisfies 
the Ginsparg-Wilson relation \cite{Ginsparg:1981bj}
and thus realizes the exact chiral symmetry
\cite{Luscher:1998pq}.
In order to avoid gauge configurations that are too rough
and the topology is not well-defined, we use the Iwasaki
action \cite{Iwasaki:1985we, Iwasaki:1984cj} combined with
the topology conserving action 
\cite{Vranas:1999rz, Izubuchi:2002pq, Fukaya:2006vs}
for the gauge part of the action.
It turned out that keeping topology is also helpful to
reduce numerical costs of the dynamical overlap fermions.
On a $16^3 \times 32$ lattice with the lattice spacing
$a\sim 0.11$--0.12~fm, we have simulated two-flavor QCD with
the quark mass down to $\sim$ 3~MeV, which is even smaller
than the physical up and down quark masses.
We found that the locality of the overlap-Dirac operator is
good enough compared to the QCD scale 
(See Ref.~[\citen{Yamada:2006fr}]).
The preliminary results from the project have been reported
in Refs.~[\citen{Fukaya:2006xp, Kaneko:2006pa, Matsufuru:2006xr}].

The outline of this article is as follows.
In Sec.~\ref{sec:chiralsym}, 
We give a brief review about how the exact chiral symmetry
and its topological properties can be realized on the lattice.
The results of numerical simulations are reported in 
Sec.~\ref{sec:lattice}.
Since our simulations are limited in a fixed topological sector
on a fixed volume lattice $\sim$ 2$^3\times 4$~fm$^4$,
the effects from finite volume and topology could be significant 
after reaching the chiral limit.
We discuss this in Sec.~\ref{sec:VQ}.
Summary and discussion are given in Sec.~\ref{sec:summary}.

\section{Chiral symmetry and topology}
\label{sec:chiralsym}
The difficulty of chiral symmetry on the lattice is
summarized by 
the Nielsen-Ninomiya theorem \cite{Nielsen:1980rz, Nielsen:1981xu}: 
any local Dirac operator which satisfies
$D\gamma_5 +\gamma_5 D=0$ must have unphysical poles.
One can easily see this in free fermions.
By the discretization, the continuum Dirac operator in the
momentum space, $D=i\gamma_\mu p_\mu$, is replaced by the
lattice counterpart
\begin{equation}
\label{eq:naivefermion}
D^{\rm naive}=\frac{\gamma_\mu}{2}(\partial_\mu + \partial^*_\mu)
\sim i\gamma_\mu \sin (ap_\mu). 
\end{equation}
It has unphysical poles at
$p_\mu = \pi/a$. 
Here, $\partial_\mu$ and $\partial^*_\mu$ denote
the forward and backward subtraction, respectively.
Wilson's prescription \cite{Wilson:1975id} to avoid the unphysical poles is to
add a higher derivative term 
\begin{equation}
\label{eq:WilsonF}
D_W = \frac{\gamma_\mu}{2}(\partial_\mu + \partial^*_\mu) 
-\frac{a}{2}\partial_\mu\partial^*_\mu  
\sim \gamma_\mu \sin (ap_\mu) + \frac{2}{a}\sum_\mu \sin^2(p_\mu a /2).
\end{equation}
This additional term, known as the Wilson term,  
gives the doublers a mass of the cut-off scale and
let them decoupled from the theory. 
but it explicitly breaks the chiral symmetry.
Since the Wilson term contains the covariant derivative,
the eigenvalue distribution of the Dirac operator 
is largely deformed from the continuum limit.
As a result, one loses the identification of the zero-modes,
and thus the clear definition of the quark mass.

Neuberger's overlap-Dirac operator 
\cite{Neuberger:1997fp,Neuberger:1998wv} 
is defined by
\begin{equation}
\label{eq:overlap}
D =\frac{1}{\bar{a}}\left(1+\gamma_5 \frac{aH_W}{\sqrt{a^2H_W^2}}\right),
\bar{a}=\frac{a}{1+s}, aH_W=\gamma_5(aD_W-1-s),
\end{equation}
where a constant $s$ is taken in a range $0< s < 1$.
It satisfies the Ginsparg-Wilson relation \cite{Ginsparg:1981bj},
$\gamma_5 D + D \gamma_5 = \bar{a}D\gamma_5 D$,
and the fermion action
$S_F=\sum_x\bar{\psi}(x)D\psi(x)$,
is exactly invariant under the chiral rotation\cite{Luscher:1998pq},
\begin{equation}
\psi \to e^{i\alpha\hat{\gamma_5}}\psi,\;\;\; 
\bar{\psi} \to e^{i\alpha \gamma_5},\;\;\;
\hat{\gamma_5}=\gamma_5(1-\bar{a}D),
\end{equation}
at finite lattice spacings.
Since $|\gamma_5 \frac{aH_W}{\sqrt{a^2H_W^2}}|^2=1$,
the eigenvalues of the overlap Dirac operator are distributed on
a circle with a radius $(1+s)/a$.
Note that this circle exactly passes the origin 
where we can count the number of the chiral zero-modes 
(all the zero-modes commute with $\gamma_5$), 
and the quark mass can be simply defined by just adding $m$ to $D$.
Because of the relation \cite{Hasenfratz:1998ri}
\begin{equation}
\label{eq:index}
n_+-n_- = \mbox{Tr} \hat{\gamma_5}/2,
\end{equation}
where $n_\pm$ denotes the number of zero-modes with $\pm$ chirality,
and the perturbative expansion of $\hat{\gamma_5}$
\cite{Kikukawa:1998pd,Adams:1998eg},
\begin{equation}
\hat{\gamma_5}(x, x)/2 = \frac{1}{32\pi^2}
\mbox{tr}\epsilon_{\mu\nu\rho\sigma}F_{\mu\nu}(x)F_{\rho\sigma}(x)+O(a^2),
\end{equation}
one can identify the index $n_+-n_-$ as the topological charge.

One should note, however, that Eq.(\ref{eq:overlap}) is ill-defined
when $H_W$ has zero-modes.
From Eq.(\ref{eq:index}),
\begin{equation}
Q\equiv n_+-n_-= \mbox{Tr} \hat{\gamma_5}/2
= \mbox{Tr} \gamma_5(1-\bar{a}D)/2
= \mbox{Tr} \gamma_5\bar{a}D/2
=\frac{1}{2}\mbox{Tr}\frac{H_W}{\sqrt{H_W^2}},
\end{equation}
one can see that the topological charge $Q$ changes when 
the eigenvalue of $H_W$ crosses zero. In other words,
the equation $H_W=0$ forms 
the topology boundaries in the configuration space.
On the topology boundaries, $H_W=0$, the overlap-Dirac operator
is not smooth and its locality is not obvious \cite{Hernandez:1998et}.
Near zero-modes of $H_W$ also cause practical problems 
that the numerical cost of approximating 
$1/\sqrt{H_W^2}$ increases as the inverse of the lowest eigenvalue
and the discontinuity of $D$ requires a huge extra numerical cost
for the dynamical quarks.

In order to achieve $H_W \neq 0$ at a finite lattice spacing 
(note that $H_W\neq 0$ is automatically satisfied in $a\to 0$ limit),
which is known as the admissibility condition
\cite{Hernandez:1998et,Luscher:1999un}, 
a transparent way is to add extra fields which produces a determinant
\cite{Vranas:1999rz, Izubuchi:2002pq, Fukaya:2006vs},
\begin{equation}
\det H_W^2/(H_W^2+m_t^2)
\end{equation}
to the theory .
It is nothing but the determinant of two-flavor Wilson fermions 
and twisted-mass ghosts with a twisted mass $m_t$.
Both of the additional fields have a cut-off scale mass and
do not affect the low-energy physics. 
With this determinant, the overlap-Dirac operator $D$ is
smooth, and its locality is guaranteed.
Furthermore, the numerical costs are largely reduced.
Since this determinant does not allow the topology change during
the hybrid Monte Carlo updates, we call it the topology
conserving action. 

\section{Lattice simulation}
\label{sec:lattice}
We generate the gauge configurations of two-flavor QCD
according to the Boltzmann weight
\begin{equation}
\exp(-\beta S_G)\det ((1-\bar{a}m/2)D+m)^2 \det H_W^2/(H_W^2+m_t^2),
\end{equation}
where $\beta S_G$ denotes the Iwasaki gauge action
\cite{Iwasaki:1985we, Iwasaki:1984cj}
with the coupling $\beta=2.3$ and 2.35, 
and  $(1-\bar{a}m/2)D+m$ is an expression of the massive
overlap fermion. 
We take $\bar{a}=a/1.6$ and $am_t=0.2$ for all the simulations.
On a $16^3\times 32$ lattice, we have performed $O(1000)$ trajectories of
the hybrid Monte Carlo updates in the $Q=0$ topological sector.
The lattice spacing is estimated to be $a\sim 0.11$-0.12~fm
from the Sommer scale $r_0$ ($=$ 0.49 fm).  
Details are seen in Refs.~[\citen{Fukaya:2006xp, Kaneko:2006pa,Matsufuru:2006xr}]

The new supercomputer system at KEK and some algorithmic improvements
\cite{Fodor:2003bh,Cundy:2004pz,Hasenbusch:2001ne}
enable us to reduce the quark mass down to $ma=0.002 (\sim$ 3~MeV), 
which is even less than the physical value. 
In fact, we observe that the numerical cost, or the number
of multiplication of $D_W$, does not depend too strongly on
the sea quark mass $m$.
However, we note that this insensitivity to $m$ indicates that
the lowest eigenvalue of $D$ is larger than the quark mass, 
due to the finite volume effects, as discussed in the next section.

\begin{figure}[tb]
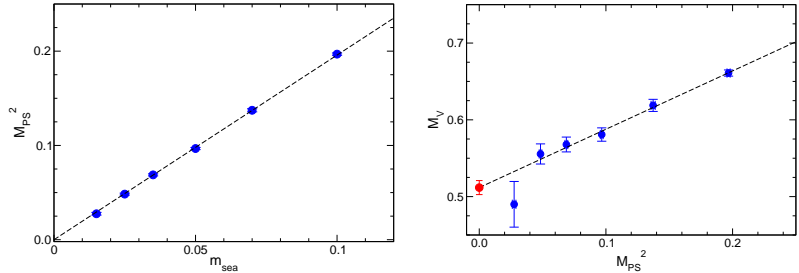

\begin{center}
   \includegraphics[angle=0,width=0.45\linewidth,clip]{mPS2_vs_mud.eps}
   \hspace{2mm}
   \includegraphics[angle=0,width=0.43\linewidth,clip]{mV_vs_mPS2.eps}
   \vspace{-3mm}
   \caption{
      Chiral extrapolation of pseudo-scalar (left) and 
      vector meson masses (right).
   }
   \label{fig:meson:chiral_fit}
   \vspace{-5mm}  
\end{center}
\end{figure}

Meson masses are evaluated from the correlation functions 
measured with smeared source and local sink operators.
Figure~\ref{fig:meson:chiral_fit} shows the chiral extrapolation of 
the pion and the vector meson masses.
The pion mass agrees well with the leading ChPT 
formula, $m_\pi^2 \propto m\Sigma /F_\pi^2$, 
where the bare quark mass $m$ needs no additive renormalization,
owing to the exact chiral symmetry.
With our present statistics, we find no significant deviation
from a simple linear fit
\begin{equation}
   m_\pi^2
    = 
   b_{\rm PS}\, m,
   \hspace{5mm}
   m_{\rm V}
   = 
   a_{\rm V} + b_{\rm V}\,m_\pi^2,
   \label{eqn:meson:chiral_fit}
\end{equation}
which gives $\chi^2/{\rm dof} \lesssim 1.0$.
With this linear fit, 
we obtain $a$ = 0.1312(23)~fm using the $\rho$ meson mass as
an input.
This is consistent with the estimate from $r_0$ within 10\% accuracy.

\section{Finite $V$ and fixed $Q$ effects}
\label{sec:VQ}.

Since our simulations are done on a fixed volume lattice 
$(\sim 1.8$-2 fm$)^4$ in a fixed topological sector,
any observable may systematic errors due to 
the finite volume and fixed topology.
Although the simplest solution to eliminate these errors is
to go to larger volumes, we can take an alternative way that
we treat finite $V$ and $Q$ effects using an effective theory.

The chiral perturbation theory (ChPT) 
\cite{Gasser:1987ah,Damgaard:2001js}
and the chiral Random Matrix theory (ChRMT) \cite{Damgaard:2000ah}
are valid in estimating the finite $V$ effects on pions,
especially when the quark mass is so small that 
the pion Compton wave length is larger than the lattice size.
This set-up is known as the $\epsilon$-regime.
In the $\epsilon$-regime, ChRMT describes the eigenvalue
spectrum of the Dirac operator with the chiral condensate
$\Sigma$ as an unique parameter.
The pion correlator in the $\epsilon$-regime 
is no more exponential but quadratic function
of $t$, of which curvature depends on the pion decay constant $F_\pi$.

\begin{figure}[tb]
\begin{center}
   \includegraphics[angle=0,width=0.45\linewidth,clip]{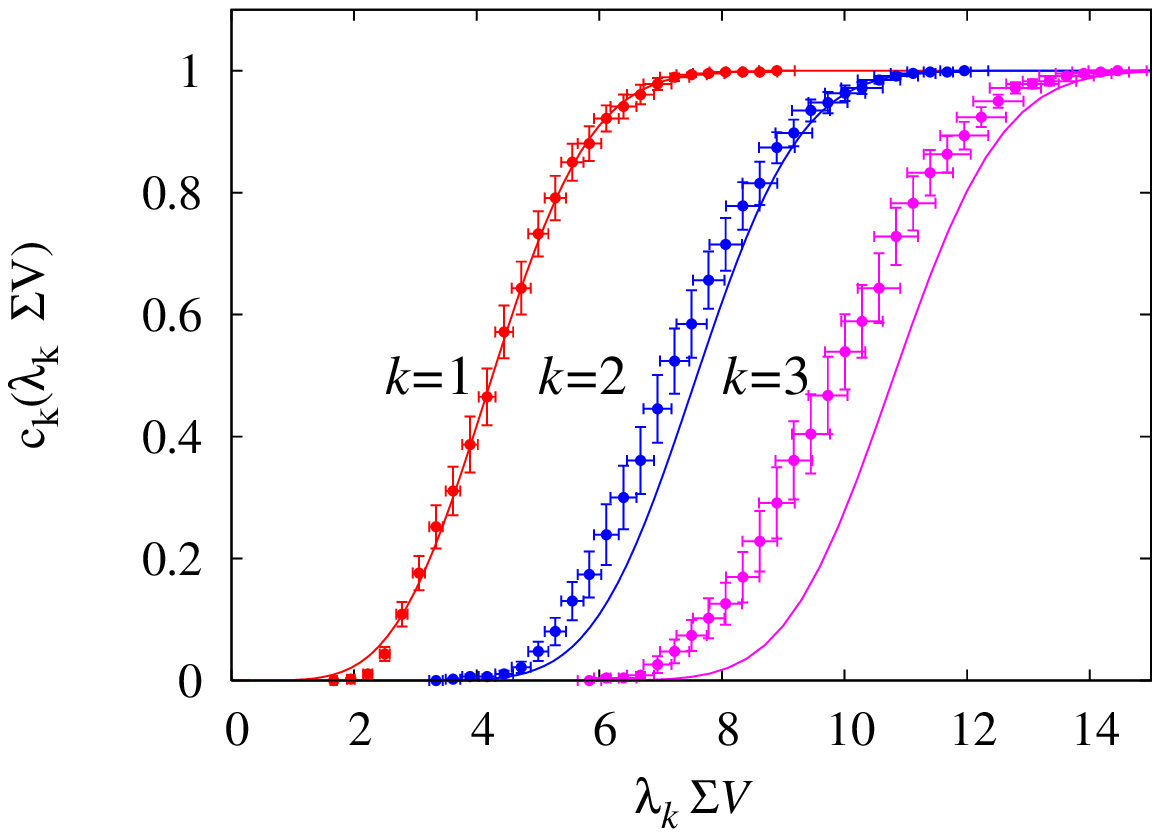}
   \hspace{2mm}
   \includegraphics[angle=0,width=0.43\linewidth,clip]{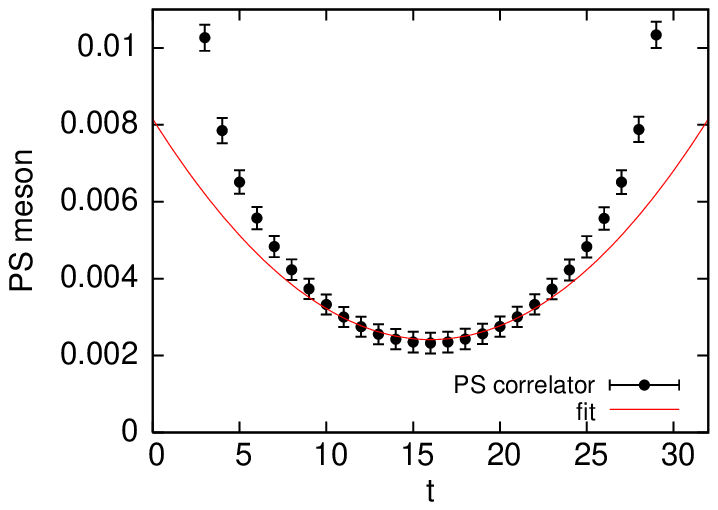}
   \vspace{-3mm}
   \caption{
      Left: the cumulative distribution of the Dirac eigenvalues.
      Right: The pion correlator.
   }
   \label{fig:e-regime}
   \vspace{-5mm}  
\end{center}
\end{figure}

We compare our numerical result at the lightest quark mass with the
prediction of ChRMT and ChPT.
The left panel of Fig.\ref{fig:e-regime} shows the cumulative
eigenvalue distribution of the overlap Dirac operator, 
from which we extract the chiral condensate as 
$\Sigma = (251(7)(11)$ MeV$)^3$,
where the errors are statistical and an estimate of the higher order
effects in the $\epsilon$-expansion.
This value is consistent with earlier works 
\cite{DeGrand:2006nv, Lang:2006ab} 
which are done with heavier quark masses.
The right panel shows the pion correlator and
the quadratic fit curve, which gives $F_\pi =86(7)$ MeV.
Note that this value is obtained near the chiral limit
without doing any chiral extrapolations.

\section{Summary and discussion}
\label{sec:summary}

The QCD simulation in the chiral regime is feasible 
with the exact chiral symmetry respected.
The topology conserving action is very helpful to reduce
the numerical cost of the dynamical overlap quarks.
We can reduce the sea quark mass to the physical up and down
quark masses or even lower.
Near the chiral limit,
the finite $V$ and fixed $Q$ effects are
important since the pion is sensitive to these effects.
Through the chiral perturbation theory or the chiral Random
Matrix Theory, these effects can even be used to extract the
low energy constants, such as the chiral condensate 
and the pion decay constant.
To extend our lattice size is, however, 
important to confirm them in the future works.


Numerical simulations are performed on Hitachi SR11000 and 
IBM System Blue Gene Solution 
at High Energy Accelerator Research Organization (KEK) 
under a support of its Large Scale Simulation Program (No.~06-13).
This work is supported in part by the Grant-in-Aid of the
Ministry of Education (No.~18840045).

\end{document}